\documentclass{article}

\evensidemargin=\oddsidemargin
\def\Title#1#2#3{%
    \baselineskip=18pt
    \begin{center}
          {\large\bf{#1} \\ }
          \bigskip\bigskip
          {#2} \\
          {#3} \\
    \end{center}}
\long\def\Abstract#1{%
         \bigskip
         \parbox{0.93\textwidth}{%
                 \begin{center}
                       {\bf Abstract} \\
                 \end{center}
                 \medskip{\baselineskip=14pt #1}
                 \vss}
         \bigskip}

\makeatletter
\renewcommand{\section}%
 {\@startsection{section}{1}{0pt}%
  {-3.25ex plus -1ex minus -.2ex}{1.5ex plus .2ex}%
  {\vspace*{5mm}\raggedright\large\bf }}
\renewcommand{\subsection}%
 {\@startsection{subsection}{2}{0pt}%
  {-2.25ex plus -.5ex minus -.2ex}{-1.5ex plus -.2ex}{\bf }}
\renewcommand{\subsubsection}%
 {\@startsection{subsubsection}{3}{0pt}%
  {-1.25ex plus -.2ex minus -.1ex}{-1.2ex plus -.2ex}{\bf }}

\makeatother

\begin{document}

\Title{Is the Wheeler -- DeWitt equation more fundamental\\
than the Schr\"odinger equation?}%
{T. P. Shestakova}%
{Department of Theoretical and Computational Physics,
Southern Federal University,\\
Sorge St. 5, Rostov-on-Don 344090, Russia \\
E-mail: {\tt shestakova@sfedu.ru}}

\Abstract{The Wheeler -- DeWitt equation was proposed 50 years ago and until now it is the cornerstone of most approaches to quantization of gravity. One can find in the literature the opinion that the Wheeler -- DeWitt equation is even more fundamental than the basic equation of quantum theory, the Schr\"odinger equation. We still should remember that we are in the situation when no observational data can confirm or reject the fundamental status of the Wheeler -- DeWitt equation, so we can give just indirect arguments in favor of or against it, grounded on mathematical consistency and physical relevance. I shall present the analysis of the situation and comparison of the standard Wheeler -- DeWitt approach with the extended phase space approach to quantization of gravity. In my analysis I suppose, firstly, that a future quantum theory of gravity must be applicable to all phenomena from the early Universe to quantum effects in strong gravitational fields, in the latter case the state of the observer (the choice of a reference frame) may appear to be significant. Secondly, I suppose that the equation for the wave function of the Universe must not be postulated but derived by means of a mathematically consistent procedure, which exists in path integral quantization. When applying this procedure to any gravitating system, one should take into account features of gravity, namely, non-trivial spacetime topology and possible absence of asymptotic states. The Schr\"odinger equation has been derived early for cosmological models with a finite number of degrees of freedom, and just recently it has been found for the spherically symmetric model which is a simplest model with an infinite number of degrees of freedom. The structure of the Schr\"odinger equation and its general solution appears to be very similar in these cases. The obtained results give grounds to say that the Schr\"odinger equation retains its fundamental meaning in constructing quantum theory of gravity.}

\section{Introduction}
50 years ago, in 1967, the seminal paper by Bryce DeWitt \cite{DeWitt} was published where the Wheeler -- DeWitt equation was presented. This equation underlies most approaches to quantization of gravity, from Quantum Geometrodynamics to Loop Quantum Gravity. At the same time, the equation suffers from shortcomings, the most known of which are the problem of time and related problems. Many reviews were devoted to the problem of time and attempts of its resolution (see, for example, \cite{Kuchar,Isham,SS1,SS2}). The criticism of the Wheeler -- DeWitt equation has led Isham \cite{Isham} to the following strong statement: ``{\it ...although it may be heretical to suggest it, the Wheeler -- DeWitt equation -- elegant though it be -- may be completely the wrong way of formulating a quantum theory of gravity}''.

At the same time, an opposite tendency has appeared in the literature, according to which the Wheeler -- DeWitt equation is more fundamental than the Schr\"odinger equation. The tendency is close related to the idea that time is something irrelevant in quantum gravity, and, perhaps, in physics in general. The irrelevance of time can be explained by the following reasoning \cite{Kiefer1}: ``{\it In classical canonical gravity, a spacetime can be represented as a `trajectory' in configuration space -- the space of all three-metrics... Since no trajectories exist anymore in quantum theory, no spacetime exists at the most fundamental, and therefore also no time coordinates to parameterize any trajectory}''. The same idea one can find in \cite{Rovelli1}: ``...{\it in quantum gravity the notion of spacetime disappears in the same manner in which the notion of trajectory disappears in the quantum theory of a particle}''.

In \cite{RV} in the Chapter titled ``Physics without time'' a simple example of a particle in some potential $U(q)$ is given that illustrates both the irrelevance of time and fundamentality of constraints. Consider a change of time variable $t=t(\tau)$ in the action,
\begin{equation}
\label{pat_act}
S=\!\int\!dt\,\left[\frac m2\left(\frac{dq}{dt}\right)^2-U(q)\right]
 =\!\int\!d\tau\,\dot t\left[\frac m2\left(\frac{\dot q}{\dot t}\right)^2-U(q)\right]
 =\!\int\!d\tau\,\left[\frac m2\frac{\dot q^2}{\dot t}-\dot t U(q)\right],
\end{equation}
where the dot denotes the derivative with respect to $\tau$. Then, one introduces the momenta
\begin{equation}
\label{pat_mom}
p_q=\frac{\partial\cal L}{\partial\dot q}=m\frac{\dot q}{\dot t};\quad
p_t=\frac{\partial\cal L}{\partial\dot t}=-\frac m2\left(\frac{\dot q}{\dot t}\right)^2-U(q).
\end{equation}
It leads to a theory with the constraint
\begin{equation}
\label{pat_con}
p_t+H(p_q,q)=0;\quad
H(p_q,q)=\frac{p_q^2}{2m}+U(q).
\end{equation}
In the operator form, after the replacement $p_t\to -i\hbar\displaystyle\frac{\partial}{\partial t}$ it gives the Schr\"odinger equation. Rovelli \cite{Rovelli2} wrote that this formalism is a generalization of standard quantum mechanics, the Wheeler -- DeWitt equation is just a generalization of the Schr\"odinger equation, to the case where a preferred time variable is not singled out. Therefore, it must be considered as more fundamental. Indeed, most fields playing a key role in modern theoretical physics are systems with constraints, most mechanical systems are not, but can be made such systems by time reparametrization. So, the constraints have a fundamental significance, and the Schr\"odinger equation can be written only in particular cases.

I agree that there does not exist a preferred time as well as a preferred reference frame. However, let me emphasize that ``a chosen time variable'' is not the same as ``a preferred time variable''. To get any solution to the Einstein equations, one has to choose some reference frame. It does not break the equality of reference frames in General Relativity. I believe that ``a chosen reference frame'' must be a bridge between quantum theory of gravity and observations which are necessary to verify it and which imply measurements in space and time. Though we are not able to do these observations now, they may be available in the future.

We should remember that we are in the situation when no observational data can confirm or reject the fundamental status of the Wheeler -- DeWitt equation, so we can give just indirect arguments in favor of or against it, grounded on mathematical consistency and physical relevance. What can we expect from quantum gravity?

It is naturally to suppose that a future quantum theory of gravity must be applicable to quantum gravitational phenomena at all scales from the early Universe to effects in strong gravitational fields. In particular, one may hope that this theory could give us a key to understanding of formation and evolution of quantum black holes as well as their final stage. It requires consideration of processes developing in time, where time is related to the state of a certain observer.

I also accept the assumption that in the future theory an equation for the wave function of the Universe must not be postulated but derived by means of a mathematically consistent procedure (at least while it is not possible to verify the equation by direct observations). Such a procedure exists in path integral quantization. When applying this procedure to any gravitating system, one should take into account features of gravity, especially, non-trivial spacetime topology. In contrast to most situations in quantum field theory, a gravitating system may not possess asymptotic states that does not enable one to be sure of gauge invariance of the path integral. This procedure leads to the Schr\"odinger equation for the wave function of the Universe, meantime the Wheeler -- DeWitt equation follows from it under some particular conditions.

Since the end of 1990s I have been developing the extended phase space approach to quantization of gravity \cite{SSV1,SSV2,SSV3,SSV4}. In the next sections I shall present the comparison of this approach with the standard Wheeler -- DeWitt quantum geometrodynamics. Section 2 is devoted to Hamiltonian dynamics in extended phase space as an alternative to generalized Hamiltonian dynamics proposed by Dirac. In Section 3 the grounds for the choice of  quantization scheme are given and the main results obtained in the extended phase space approach are presented, in particular, the general structure of solutions to the Schr\"odinger equation is discussed. Section 4 comprises the interpretation of these results and conclusions.

\section{Hamiltonian dynamics in extended phase space}
In 1950--1958 Dirac formulated his ``generalized Hamiltonian dynamics'' \cite{Dirac1,Dirac2}. His main aim was to elaborate a Hamiltonian form of a theory with constraints as the first step towards its further quantization. He wrote in \cite{Dirac3}: ``{\it Any dynamical theory must first be put in the Hamiltonian form before one can quantize it}''. Ironically, the generalized Hamiltonian dynamics played no role in the creation of very successful and experimentally verified gauge theories like quantum electrodynamics. I would point out three features of the Dirac approach: (i) All variables are divided into physical and non-physical ones, only the former ones being included into (physical) phase space; (ii) a linear combination of constraints must be added to the Hamiltonian constructed by usual rules from physical variables; (iii) after quantization constraints become conditions on a state vector.

These three statements can be considered as postulates, since their necessity has never been proved. The confidence in these statments is based rather on the authority of Dirac than on the efficiency of any theory founded on these postulates. Strictly speaking, the only theory which rests upon them is the Wheeler -- DeWitt quantum geometrodynamics.

In 1970s a new approach to quantization of gauge theories based on path integration was proposed by Batalin, Fradkin and Vilkovisky \cite{BFV1,BFV2,BFV3}. In fact, they rejected the statement (i) and included physical, non-physical (gauge) and ghost variables into extended phase space. However, in their approach gauge variables were still considered as non-physical, secondary degrees of freedom playing just an auxiliary role in the theory. The Hamiltonian form of action was constructed in such a way that the Hamiltonian coincides with the one built by the prescription of Dirac. There were attempts to prove the equivalence between the Batalin -- Fradkin -- Vilkovisky (BFV) and Dirac approaches. In particular, in \cite{Hennaux} the proof was given for the case when constraints commute. The assumption was made that any system of constraints can be reduced to an equivalent system of commuting constraints, though it may not be practically easy to find the equivalent commuting constraints. Also, the ordering problem was ignored. It is hardly possible to give a proof for an arbitrary system of constraints, let alone taking into account the ordering problem.

On the other hand, there exist another way to construct Hamiltonian dynamics of a constrained system exploiting the idea of extended phase space. The main source of difficulties with the Hamiltonian formulation was the impossibility to construct the Hamiltonian according to the usual rule
$H=p\dot q-L$, since for gauge variables their generalized velocities are missing in the Lagrangian and could not be expressed in coordinates and momenta. But the notion of extended phase space came from the path integral approach, where the gauge invariant action of the original theory is replaced by an effective action that includes gauge fixing and ghost terms. For a system with a finite number of degrees of freedom a general enough form of the action can look like
\begin{equation}
\label{action}
S=\!\int\!dt\,\left[\displaystyle\frac12 g_{ab}(N, q)\dot q^a\dot q^b-U(N, q)
  +\pi\left(\dot N-\frac{\partial f}{\partial q^a}\dot q^a\right)
  +N\dot{\bar\theta}\dot\theta\right].
\end{equation}
Here $q=\{q^a\}$ stands for physical variables and $N$ denotes a gauge variable (it may be, for example, the lapse function), $\theta$, $\bar\theta$ are the Faddeev -- Popov ghosts. The gauge condition for $N$,
\begin{equation}
\label{gauge}
N=f(q)+k;\quad
k={\rm const},
\end{equation}
can be presented in a differential form,
\begin{equation}
\label{diff_gauge}
\dot N=\frac{\partial f}{\partial q^a}\dot q^a.
\end{equation}
The gauge condition (\ref{diff_gauge}) introduces into the effective Lagrangian the missing velocity $\dot N$, so that the Lagrange multiplier $\pi$ of the gauge condition plays the role of the momentum conjugate to $N$. Now the Hamiltonian can be constructed by the rule
\begin{equation}
\label{Ham1}
H=p_a\dot q^a+\pi\dot N+\bar{\cal P}\dot\theta+\dot{\bar\theta}{\cal P}-L.
\end{equation}
$\bar{\cal P}$, $\cal P$ are ghost momenta. The terms with $N$ are reduced in (\ref{Ham1}), so we shall come to the expression
\begin{eqnarray}
\label{Ham2}
H&=&\frac12g^{ab}p_a p_b+\pi p_a f^{,a}+\frac12\pi^2 f_{,a}f^{,a}-U(N, q)+\frac1N\bar{\cal P}{\cal P}\nonumber\\
&=&\frac12G^{\alpha\beta}P_{\alpha}P_{\beta}+U(N, q)+\frac1N\bar{\cal P}{\cal P},
\end{eqnarray}
where
\begin{equation}
\label{Galphabeta}
f_{,a}=\frac{\partial f}{\partial q^a};\quad
G^{\alpha\beta}=\left(
\begin{array}{cc}
f_{,a}f^{,a}&f^{,a}\\
f^{,a}&g^{ab}
\end{array}
\right);\quad
Q^{\alpha}=(N,\,q^a);\quad
P_{\alpha}=(\pi,\,p_a).
\end{equation}

Hamiltonian dynamics in this form resembles that of an unconstrained system to the large extent. Hamiltonian equations are fully equivalent to the Lagrangian set of equations obtained from the effective action (\ref{action}) by variational procedure. The extended Lagrangian set of equations contains motion equations, the constraint, the gauge condition and ghost equations. All of them (including the constraint and the gauge condition) have a status of Hamiltonian equations in extended phase space. One can object that the Lagrangian equations, as well as the Hamiltonian equations in extended phase space, differ from equations of the original gauge invariant theory by terms resulting from variation of gauge fixing and ghost parts of the action. But, if one speaks about General Relativity, one must remember that any solution to the gauge-invariant Einstein equations is determined up to arbitrary functions which have to be fixed by a choice of a reference frame. It is usually done at the final stage of solving the Einstein equations. In the present approach we introduce the gauge fixing function $f(q)$ from the very beginning, though one can keep this function  non-fixed up to the final stage of solving the equations. The formalism works for any gauge condition; no gauge condition is privileged. {\it It is important that one cannot avoid fixing a reference frame to obtain a final form of the solution}. Under the conditions $\pi=0$, $\theta=0$, $\bar\theta=0$ all gauge-dependent terms are excluded from the equations and they are reduced to gauge-invariant ones; therefore, any solution of the Einstein equations can be found among solutions of the extended set.

The equivalence of the extended Lagrangian set of equations and the Hamiltonian equations in extended phase space was demonstrated for the closed isotropic cosmological model in \cite{Shest1}, for models with finite number of degrees of freedom -- in \cite{SSV4}, for the spherically symmetric gravitational model -- in \cite{Shest2}. Let me emphasize that the equivalence can be demonstrated for arbitrary parametrization of gravitational variables.

The equivalence of the Lagrangian and Hamiltonian formulations is not a trivial problem. Among many approaches to quantization of gauge theories the more powerful are the BFV approach based on the Hamiltonian formulation and the Batalin -- Vilkovisky (BV) approach \cite{BV} based on the Lagrangian formulation, the latter is known also as the antibracket -- antifields formalism. The both are applicable to any gauge theory, including theories with open algebras. For gravity, the effective action in the BV formalism is reduced to the Faddeev -- Popov action, and for a system with a finite number of degrees of freedom it has the form like (\ref{action}). The equivalence of the BFV and BV approaches was studied, for example, in \cite{FH}. However, the starting point in this paper was the Hamiltonian formalism. The authors have not considered the Lagrangian form of the action similar to (\ref{action}), so their proof of the equivalence does not seem to be general enough. Among other works let us mention the proof of the perturbative equivalence between the BFV and BV quantization schemes given in \cite{BT}. In the proposed approach the demonstration of the equivalence between the Lagrangian and Hamiltonian formulations is straightforward and can be checked by direct calculations.

In the Dirac approach constraints play the role of generators of transformations in phase space. They produce correct transformation only for physical variables. By {\it correct} transformations I mean the ones which coincide with gauge transformations in the Lagrangian formalism. One cannot require that the constraints must produce correct transformations for gauge variables since they are considered to be redundant in the Dirac scheme. However, in the extended phase space formulation we would like to have a generator that would produce correct transformations for all degrees of freedom. In the BFV approach the generator is the BRST charge which can be constructed as a series in Grassmannian (ghost) variables with coefficients given by generalized structure functions of constraints algebra \cite{Hennaux}:
\begin{equation}
\label{BFV_BRST}
\Omega_{BFV}=c^{\alpha}U^{(0)}_{\alpha}+c^{\beta}c^{\gamma}U^{(1)\alpha}_{\gamma\beta}\bar\rho_{\alpha}+\ldots,
\end{equation}
$c^{\alpha}$, $\bar\rho_{\alpha}$ are the BFV ghosts and their conjugate momenta, $U^{(n)}$ are $n$th order structure functions, while zero order structure functions $U^{(0)}_{\alpha}=G_{\alpha}$ being Dirac constraints. The BFV effective action is
\begin{equation}
\label{S_BFV}
S_{BFV}=\!\int\!dt\,\left[p_a\dot q^a+\pi\dot N+\dot c^{\alpha}\bar\rho_{\alpha}+\{\bar\psi,\Omega_{BFV}\}\right].
\end{equation}
The bracket $\{\bar\psi,\Omega_{BFV}\}$ with the BRST charge (\ref{BFV_BRST}) and under special choice of $\bar\psi$ leads to the Hamiltonian which, as has been already mentioned, coincides with the one constructed by the prescription of Dirac. The BRST charge, as well as the constraints, do not produce correct transformations for gauge variables which are on an equitable basis with physical variables now. As the authors of the BFV approach emphasized themselves \cite{BFV1}, in the gravitational theory the gauge transformations cannot be presented as canonical transformations, and thus they differ from transformations generated by constraints. Actually, the BV and BFV formulations are {\it two non-equivalent theories with different groups of transformations}.

Again, there exist an alternative way to construct the BRST charge making use of global BRST invariance and the first Noether theorem. It is presented in \cite{SSV4} for models with a finite number of degrees of freedom, and in \cite{Shest2} for the spherically symmetric model. For the model with the action (\ref{action}) it looks like
\begin{equation}
\label{BRST_mod}
\Omega=-H\theta-\pi{\cal P},
\end{equation}
$H$ is the Hamiltonian in extended phase space (\ref{Ham2}). It is very important that to get correct transformations one should use the Hamiltonian equations in extended phase space and their equivalence with the Lagrangian set of equations. In particular, to get the transformation for the gauge variable $N$ one should use the equation $\dot N=\displaystyle\frac{\partial H}{\partial\pi}$ which is, in fact, equivalent to the gauge condition (\ref{diff_gauge}).

Historically, different authors used various parametrizations of gravitational variables. Dirac \cite{Dirac3} dealt with original variables, which are components of the metric tensor, whereas the most famous parametrization is probably that of Arnowitt -- Deser -- Misner (ADM) \cite{ADM} in terms of the lapse and shift functions. From the point of view of the Lagrangian formalism, all the parametrizations are rightful, and the correspondent formulations are equivalent. However, if a parametrization touches upon gauge degrees of freedom, the relation between old and new gravitational variables may not be a canonical transformation. It cannot be a canonical transformation in the Dirac approach where gauge degrees of freedom are not canonical variables. But, even if one formally extends phase space by including gauge variables into it, it may be not enough. The example is given in \cite{KK}, where the transformation from components of the metric tensor to the ADM variables is considered,
\begin{equation}
\label{ADM-tr}
g_{00}=\gamma_{ij}N^i N^j-N^2,\qquad
g_{0i}=\gamma_{ij}N^j,\qquad
g_{ij}=\gamma_{ij}.
\end{equation}
Then the Poisson bracket $\{N,\,\Pi^{ij}\}$ between the lapse function $N$ and the momenta $\Pi^{ij}$ conjugate to $\gamma_{ij}$ is calculated and it appears to be non-zero. It implies that the Dirac Hamiltonian formulation for gravitation and the ADM one are not equivalent from the point of view of the canonical formalism though the both are believed to be equivalent to the original (Lagrangian) formulation of General Relativity.

In \cite{Shest1} it has been shown that this contradiction can be resolved if one introduces into the Lagrangian missing velocities corresponding to gauge variables by means of differential gauge conditions, that not just formally but {\it actually} extends phase space. It was demonstrated for the full gravitational theory for the class of transformations
\begin{equation}
\label{new-var}
g_{0\mu}=v_{\mu}\left(N_{\nu},\gamma_{ij}\right),\qquad
g_{ij}=\gamma_{ij}
\end{equation}
from components of the metric tensor to new variables $N_{\mu}$, $\gamma_{ij}$ (the transformation (\ref{ADM-tr}) is a particular case of (\ref{new-var})). A general form of gauge conditions, $f^{\mu }(g_{\nu\lambda})=0$, was used, the differential form of which is
\begin{equation}
\label{diff-g}
\frac{d}{dt}f^{\mu}(g_{\nu\lambda})=0,\qquad
\frac{\partial f^{\mu}}{\partial g_{00}}\dot g_{00}
 +2\frac{\partial f^{\mu}}{\partial g_{0i}}\dot g_{0i}
 +\frac{\partial f^{\mu}}{\partial g_{ij}}\dot g_{ij}=0.
\end{equation}

The canonicity of the transformation (\ref{new-var}) has been proved, thus one can expect that Hamiltonian formulations in extended phase space must be equivalent for various parametrizations from this class.

So, the proposed formulation of Hamiltonian dynamics is self-consistent, it removes some shortcomings of the Dirac generalized dynamics: it is fully equivalent to the Lagrangian formulation; the generator of transformations in extended phase space constructed in accordance with the Noether theorem produces correct gauge transformations for all gravitational degrees of freedom; introducing new gravitational variables from the class of parametrizations (\ref{new-var}) is proved to be a canonical transformation.

\section{Quantization scheme: why it leads to the Schr\"odinger equation?}
The simplest way to get the Wheeler -- DeWitt equation is to replace components of the metric tensor and their conjugate momenta by operators in the so-called Hamiltonian gravitational constraint in the manner described in \cite{Rovelli2}. So, one comes to the famous equation
\begin{equation}
\label{WDW}
\left[\frac1{2\sqrt\gamma}\left(\gamma_{ik}\gamma_{jl}+\gamma_{il}\gamma_{jk}-\gamma_{ij}\gamma_{kl}\right)
 \frac{\delta}{\delta\gamma_{ij}}\frac{\delta}{\delta\gamma_{kl}}+\sqrt{\gamma}R^{(3)}\right]\Psi=0,
\end{equation}
$\gamma_{ij}$ are space components of the metric tensor, $\gamma$ is the determinant of $\gamma_{ij}$,  $R^{(3)}$ is 3-curvature. Eq.(\ref{WDW}) is a realization of the Dirac postulate according to which after quantization constraints become conditions on a state vector. So, the Wheeler -- DeWitt equation can also be considered as a postulate. DeWitt followed the way how Schr\"odinger had got his wave equation, and one may object that the Schr\"odinger equation was written by its author without rigorous grounds as well. However, the predictions made by means of this equation immediately confirmed its fundamental status.

Three other gravitational constraints after replacement momenta by operators take the form
\begin{equation}
\label{mom-constr}
D_j\left(\frac{\delta\Psi}{\delta\gamma_{ij}}\right)=0,
\end{equation}
$D_i$ denotes covariant derivative in three dimensions.

It is generally accepted that the constraints as conditions on a state vector ensure gauge invariance of the latter. It has been shown (see, for example, \cite{Kiefer2}) that the wave functional is invariant under three-dimensional infinitesimal diffeomorphisms assuming that infinitesimal parameters vanish at infinity. The situation with four-dimensional diffeomorphisms is more complicated. The quantum constraints (\ref{WDW}), (\ref{mom-constr}) are written in the form not depending on gauge variables thanks to the choice of the ADM parametrization. However, Hawking and Page \cite{HP} pointed out that the DeWitt supermetric on space of all three-dimensional metrics $\gamma_{ij}$ does depend on the lapse function $N$. In principle, it gives rise to a family of the Wheeler -- DeWitt equations corresponding to different relations between $N$ and $\gamma_{ij}$. Hawking and Page proposed to regard $N$ as a field independent on $\gamma_{ij}$. It is in correspondence with the choice made by DeWitt \cite{DeWitt}: $N=1$, $N_i=0$. But these additional conditions on the lapse and shift functions implicitly fix a reference frame. The same idea can be expressed in the other words: the ADM parametrization introduces in 4-dimensional spacetime $(3+1)$-splitting that is equivalent to a choice of a reference frame \cite{MM}. It raises doubts if the theory based upon the Wheeler -- DeWitt equation is indeed gauge invariant.

We saw in the previous section that there is another possibility to formulate Hamiltonian dynamics of a constrained system but the method proposed by Dirac. Hamiltonian equations in extended phase space include a constraint, however, this constraint is modified in comparison with the Hamiltonian constraint in the Dirac formalism. If one accepts the Dirac postulate about constraints after quantization as conditions on a state vector, what form of the constraint should be chosen? Should we consider the ADM parametrization as a privileged one and bear in mind the DeWitt choice $N=1$, $N_i=0$ although it implicitly fix a reference frame, instead of using the formalism in which all parametrizations are rightful?

Another way to come to the Wheeler -- DeWitt equation is to require BRST invariance of physical states. In the case of commuting constraints one gets from (\ref{BFV_BRST})
\begin{equation}
\label{phys-states}
\hat\Omega_{BFV}\Psi=0\quad\Longrightarrow\quad
\hat G_{\alpha}\Psi=0.
\end{equation}
So, for gravity the quantum constraints (\ref{WDW}), (\ref{mom-constr}) result from the requirement of BRST invariance. However, as we saw in the previous section, the BRST charge which generates correct gauge transformations for all degrees of freedom can be constructed according to the Noether theorem, and its structure does not coincide with that of the BFV generator (\ref{BFV_BRST}). In particular, the charge (\ref{BRST_mod}) does not lead to the Wheeler -- DeWitt equation. And again, one can pose the question, if it is correct to construct Hamiltonian dynamics in extended phase space so that it mimics the Dirac generalized dynamics?

There were some works devoted to derivation of the Wheeler -- DeWitt equation from a path integral. Barvinsky and Ponomariov \cite{BP} used as a starting point a path integral over the so-called reduced phase space. Halliwell \cite{Hall} relied upon the BFV quantization scheme. All the mentioned authors made use of asymptotic boundary conditions for ghosts and Lagrange multipliers of gauge fixing terms. Asymptotic boundary conditions came from ordinary quantum field theory where one usually considers systems with asymptotic states in which physical and non-physical degrees of freedom could be separated from each other. The boundary conditions ensure gauge invariance of the path integral, as well as of a theory as a whole. So, the work by Halliwell \cite{Hall} is based upon the Fradkin -- Vilkovisky theorem which states that the path integral is independent of the choice of gauge fixing function under asymptotic boundary conditions. It let him choose the simplest gauge condition $\dot N=0$ (which is in accordance with the DeWitt condition $N=1$), and that is why one cannot find any vestiges of gauge fixing function in the equation he obtained.

But the situation in the theory of gravity differs from that in quantum field theory. The only case of a gravitating system with asymptotic states is the case of asymptotically flat spacetime. A universe with non-trivial topology, in particular, a closed universe, does not possess asymptotic states. Asymptotic boundary conditions are not justified for quantum gravity. This fact should be taken into account when deriving the Wheeler -- DeWitt equation from a path integral. However, in the absence of the boundary conditions gauge invariance breaks down. The quantum version of the Hamiltonian constraint loses its sense.

In this situation I give preference to the Schr\"odinger equation as a fundamental equation which maintains its validity in quantum field theory. In accordance with the assumption made in Section 1, the equation for the wave function of the Universe must not be postulated but derived from a path integral by means of a mathematically consistent procedure. The path integral formalism does not require to construct the Hamiltonian form of the theory at all since the equation for the wave function can be derived directly from a path integral in the Lagrangian form. The Lagrangian formalism corresponds to the original (Einstein) formulation of the gravitational theory. In the extended phase space approach we deal with the path integral with the Batalin -- Vilkovisky effective action which, as was mentioned above, is reduced to the Faddeev -- Popov effective action for gravity. We consider the path integral {\it without} asymptotic boundary conditions. The generalization of the standard method originated from Feynman \cite{Feyn,Cheng} results in the following Schr\"odinger equation:
\begin{equation}
\label{SE1}
i\,\frac{\partial\Psi(N,q,\theta,\bar\theta;\,t)}{\partial t}
 =H\Psi(N,\,q,\,\theta,\,\bar\theta;\,t),
\end{equation}
where
\begin{equation}
\label{H}
H=-\frac1N\frac{\partial}{\partial\theta}
   \frac{\partial}{\partial\bar\theta}
  -\frac1{2M}\frac{\partial}{\partial Q^{\alpha}}MG^{\alpha\beta}
   \frac{\partial}{\partial Q^{\beta}}+U(N, q)-V[f];
\end{equation}
the operator $H$ corresponds to the Hamiltonian in extended phase space (\ref{Ham2}). This is another argument in favor of our choice of quantization scheme. $M$ is the measure in the path integral, $V[f]$ is a quantum correction to the potential $U$, the analogue of the term with the scalar curvature of configurational space in \cite{Cheng}. The explicit form of (\ref{H}) for an arbitrary parametrization of the gauge variable is given in \cite{Shest3}.

The wave function is defined on extended configurational space with the coordinates $N,\,q,\,\theta,\,\bar\theta$. The general solution to the Schr\"odinger equation has the following structure:
\begin{equation}
\label{GS1}
\Psi(N,\,q,\,\theta,\,\bar\theta;\,t)
 =\int\Psi_k(q,\,t)\,\delta(N-f(q)-k)\,(\bar\theta+i\theta)\,dk.
\end{equation}
It is a superposition of eigenstates of a gauge operator,
\begin{equation}
\label{k-vector}
\left(N-f(q)\right)|k\rangle=k\,|k\rangle;\quad
|k\rangle=\delta\left(N-f(q)-k\right).
\end{equation}
The function $\Psi_k(q,\,t)$ describes a state of the physical subsystem for a reference frame fixed by the condition (\ref{gauge}). It is a solution to the equation
\begin{equation}
\label{phys.SE}
i\,\frac{\partial\Psi_k(q;\,t)}{\partial t}
 =H_{(phys)}[f]\Psi_k(q;\,t).
\end{equation}
This equation can be called the Schr\"odinger equation for the physical part of the wave function, while $H_{(phys)}[f]$ can be called the physical Hamiltonian operator. It can be obtained from (\ref{H}) by separating ghosts and substituting the gauge condition (\ref{gauge}) into (\ref{H}):
\begin{equation}
\label{phys.H}
H_{(phys)}[f]=\left.\left(-\frac1{2M}\frac{\partial}{\partial q^a}M g^{ab}\frac{\partial}{\partial q^b}
 +U(N, q)-V[f]\right)\right|_{N=f(q)+k}.
\end{equation}

Recently the Schr\"odinger equation has been derived for the spherically symmetric model which is a simplest model with an infinite number of degrees of freedom. The interval for the model is
\begin{equation}
\label{spheric}
ds^2=-N^2(t,r)dt^2+V^2(t,r)dr^2+W^2(t,r)(d\theta^2+\sin^2\theta d\varphi^2).
\end{equation}
The details of derivation of the Schr\"odinger equation will be published elsewhere. I would like only mention that the structure of its general solution repeats (\ref{GS1}):
\begin{equation}
\label{GS2}
\Psi[N,\,V,\,W,\,\theta,\,\bar\theta;\,t]
 =\int\Phi_k[V,\,W,\,t]\,\delta(N-F(V,\,W)-k)\,(\bar\theta+i\theta)\,dk.
\end{equation}
$\Phi_k[V,\,W,\,t]$ is the physical part of the wave functional $\Psi[N,\,V,\,W,\,\theta,\,\bar\theta;\,t]$. Just as $\Psi_k(q,\,t)$, it is a solution to the equation similar to (\ref{phys.SE}), whose form depends on the gauge condition $N=F(V,W)+k$.

Can we come to the Wheeler -- DeWitt equation starting from the Schr\"odinger equation? Yes, we can do it if we (i) put $E=0$ in stationary solutions to the Schr\"odinger equation for the physical part of the wave function that is equivalent to rejecting time evolution of the wave function, (ii) choose the ADM parametrization and (iii) choose the gauge conditions $N=1$ $N_i=0$. In this case Eq.(\ref{phys.SE}) would reduce to the Wheeler -- DeWitt equation
\begin{equation}
\label{WDW.eq}
H_{WDW}\Psi(q)=0.
\end{equation}
with the Hamiltonian operator
\begin{equation}
\label{WDW_Ham}
H_{WDW}=-\frac1{2M}\frac{\partial}{\partial q^a}M g^{ab}\frac{\partial}{\partial q^b}+U(q).
\end{equation}

So, from the point of view of the extended space approach, the Wheeler -- DeWitt equation and its solutions answer just the particular case, namely, the particular choice of parametrization and gauge conditions and the zero eigenvalue of the Hamiltonian. If one declare that the Wheeler -- DeWitt equation is more fundamental, it means rejecting all other cases and all other solutions to the Schr\"odinger equation (\ref{SE1}). At the present level of development of the theory it seems to be bad-grounded to give preference to the Wheeler -- DeWitt equation excluding all other possibilities.

\section{Interpretation and conclusions}
As we can see, the physical picture one can get in the proposed approach strongly depends on a chosen reference frame. Is it a bad news for this approach? The answer ought to be ``Yes'', if one postulates that quantum theory of gravity must be gauge invariant. However, we do not know what this theory will be.

Let us return to the general solution (\ref{GS1}). It can be interpreted in the spirit of Everett's ``relative state'' formulation. To see this, let us appeal to the seminal paper by Everett \cite{Everett}. His first goal was to make quantum theory applicable to isolated systems containing observers in interaction with other subsystems that leads to the concept of ``relativity of states''. This concept was illustrated with a toy model originated from von Neumann \cite{Neumann}. The feature of this consideration is the existence of a relation between parameters of a quantum object (such as the position of a particle, $x$) and those of a measuring device (such as the position of a pointer, $r$). For example, in the situation when the pointer coordinate $r$ is definite, the wave function of the total system can be written as
\begin{equation}
\label{toymod}
\psi_T^{S+A}=\int\frac1{N_{r'}}\;\xi^{r'}\!(x)\;\delta(r-r')\;dr',
\end{equation}
where ${\xi^{r'}\!(x)}$ are the relative states for the measuring device states $\delta(r-r')$ of definite value $r=r'$, $N_{r'}$ is a normalization constant (see the last equation on p.~456 in \cite{Everett}).

Let us compare the expressions (\ref{toymod}) and (\ref{GS1}). Each element of the superposition (\ref{toymod}) describes a state in which the only degree of freedom of the measuring device, namely, the pointer position $r$, is definite. In General Relativity the observer is represented by the reference frame only, by means of which he is able to judge about spacetime geometry. In other words, the reference frame plays the role of a measuring device in General Relativity. Each element of the superposition (\ref{GS1}) describe a state in which the only gauge degree of freedom $N$ is definite, so that time scale is determined by processes in the physical subsystem through the gauge fixing function $f(q)$. The superposition (\ref{GS1}) is a packet over $k$, and, though a form of the packet is not determined by Eq.(\ref{phys.SE}), it has to be sufficiently narrow for $\Psi_k(q,\,t)$ to be a normalizable function \cite{SSV4}. The spread of $k$ reflects the fact that in quantum gravity the reference frame cannot be fixed absolutely precisely, in particular, time intervals between two hypersurfaces cannot be measured with arbitrary accuracy. One can say that the function $\Psi_k(q,\,t)$ describes a relative state of the physical subsystem for the reference frame fixed by the condition (\ref{gauge}). So, the extended phase space approach to quantization of gravity can be considered as a mathematical realization of the Everett concept of relative states.

This approach may have some advantages for description of quantum effects in strong gravitational fields. It is well known that even in classical General Relativity different observers (the observer who crosses a black hole horizon and the one at some distance from the black hole) watch different physical phenomena. The same is true if one considers quantum field theory in curved background. One can say that quantum states observed by various observers live in different Hilbert spaces. Susskind and his collaborators \cite{Suss1,Suss2} formulated this idea as the complementarity principle for black holes: the observer falling into a black hole and the distant observer will see physical phenomena which are complementary to each other. This principle combines two kinds of complementarity: the complementarity of physical pictures observed in different reference frames inherited from General Relativity and the quantum complementarity in the sense of Bohr. One can expect that this principle will be reflected in the full quantum gravity.

The Wheeler -- DeWitt approach does not admit the description of phenomena related to different reference systems. The description of them, as well as the appearance of time, is possible only in quasiclassical approximation (see, for example, \cite{Kiefer3}). Just the opposite, in the extended phase space approach, which leads to the time-dependent Schr\"odinger equation and where any quantum state is related to a chosen reference frame, there exist a theoretical possibility to give a complete consideration of these phenomena.

In conclusion I would like to repeat that at present we do not know what approach is correct and we would not know it until observation data do not give us a hint. However, it is too early to deprive the Schr\"odinger equation of its fundamental status.

\section*{Acknowledgments}
I am grateful to the organizers of the 3rd International Conference on Particle Physics and Astrophysics for invitation to give a talk and hospitality at the conference.

\end{document}